\documentclass[a4paper,twocolumn,amsfonts,amssymb,amsmath]{revtex4}%
\usepackage{graphicx}
\usepackage[thinspace,thinqspace,amssymb,textstyle]{SIunits}
\usepackage{braket}
\usepackage{hyperref}
\usepackage[all]{hypcap}
\usepackage{bm}

\renewcommand{\vec}[1]{\bm{#1}}

\begin{document}

\title{$d$-wave Superfluid with Gapless Edges in a Cold Atom Trap}
\author{Anne-Louise Gadsb\o lle$^{1,2}$}
\author{H. Francis Song$^{1}$}
\author{Karyn Le Hur$^{1,3}$}

\affiliation{$^1$ Department of Physics, Yale University, New Haven, Connecticut 06520, USA}
\affiliation{$^2$ Lundbeck Foundation Theoretical Center for Quantum System Research,
Department of Physics and Astronomy, Aarhus University, DK-8000 Aarhus C, Denmark}
\affiliation{$^3$ Centre de Physique Th\' eorique, \' Ecole Polytechnique, CNRS, 91128 Palaiseau Cedex, France}
\date{\today}

\begin{abstract}
We consider a strongly repulsive fermionic gas in a two-dimensional optical lattice confined by a harmonic trapping potential. To address the strongly repulsive regime, we consider the $t-J$ Hamiltonian. The presence of the harmonic trapping potential enables the stabilization of coexisting and competing phases. In particular, at low temperatures, this allows the realization of a $d$-wave superfluid region surrounded by purely (gapless) normal edges. Solving the Bogoliubov-de Gennes equations and comparing with the local density approximation, we show that the proximity to the Mott insulator is revealed by a downturn of the Fermi liquid order parameter at the center of the trap where the $d$-wave gap has a maximum. The density profile evolves linearly with distance. 
\end{abstract}

\maketitle

Ultracold atoms in optical lattices are ideal quantum simulators of complex many-body Hamiltonians that arise in condensed-matter systems \cite{Zoller,Bloch,Georges}. They embody very clean systems which can be tuned in a very precise and controlled manner from the weak to the strong coupling limit. This enables one to investigate a plethora of interesting phenomena ranging from the dynamics of strongly correlated bosons \cite{Greiner,Trotzky} and fermions \cite{Kohl,Jin} to quantum magnetism \cite{Greiner2}. The pace of experimental progress is quite impressive. Bunching and antibunching effects in the density-density correlations were found for bosons and fermions \cite{Folling}. In particular, the fermionic Hubbard model has been realized for repulsive and attractive interactions. Fingerprints of the Mott state \cite{Esslinger,Bloch2} and $s$-wave superfluidity \cite{Wolfgang} have already been observed in experiments. The challenge remains to access the N\' eel phase \cite{Jordens} in order to reveal the presence of high-temperature $d$-wave superfluidity close to half-filling \cite{Hofstetter}. Spin fluctuations, which are predicted to be the glue for $d$-wave superfluidity in the repulsive Hubbard model \cite{Anderson,Lee,Ogata,Karyn}, have been studied experimentally in the context of BEC-BCS crossover \cite{Wolfgang2}. In this Letter, we theoretically address the effect of the harmonic potential, which originates from the Gaussian profile of the laser beams generating the trap, on the $d$-wave superfluid phase of the Hubbard model. We explore the strongly repulsive limit which is realized in high-$T_{c}$ superconductors \cite{Anderson,Lee,Ogata,Karyn}.

We consider relatively small fillings such that only a superfluid cloud is left in the center of the trap and antiferromagnetism is hindered by the motion of atoms \cite{Andersen}. We investigate the coexistence between $d$-wave superfluidity and a normal phase at the boundaries applying an effective theory of a doped Mott insulator. More precisely, we start from the $t-J$ Hamiltonian and apply a low-energy (superfluid) theory that allows us to describe the proximity to the Mott insulator \cite{ZGR,Anderson2,Karyn2}. 

\begin{align}  \label{Hamiltonianstart}
\hat{H}&=-t\sum_{\braket{ij} \sigma}\left[ \hat{c}_{i \sigma}^{\dagger}\hat{c}_{j \sigma}+h.c.\right]+\sum_{i \sigma} V_{i} \hat{c}_{i \sigma}^{\dagger}\hat{c}_{i \sigma} + J \sum_{\braket{ij}} \vec{S}_{i} \cdot \vec{S}_{j},
\end{align}
where $\hat{c}_{i\sigma}^{\dagger}$ creates a particle with spin $\sigma$ at the $i$'th lattice site, $t$, the tunneling amplitude, is chosen to be spin- and direction-independent, $V_{i}$ denotes an isotropic and spin-independent external trapping potential, 
$V(x,y)=(m/2)\omega^2(x^2+y^2)$ evaluated at the $i$'th lattice site, and $\braket{ij}$ denotes nearest neighbor pairs. The last term describes the super-exchange process where $\vec{S}_{i}=1/2\sum_{\sigma \sigma'}\hat{c}_{i\sigma}^{\dagger}\sigma_{\sigma \sigma'}^{i}\hat{c}_{i \sigma'}$ is the spin-operator and $J=4t^2/U$. The number of fermions per site is assumed to be less than one. The ground state properties can be analyzed using a projected Bardeen-Cooper-Schrieffer (BCS) state taking into account that double occupancy is forbidden for large interactions, 
$\hat{P}_{G}\ket{\rm{BCS}}$ where $\hat{P}_{G}=\prod_{i}\left(1-\hat{n}_{i \uparrow}\hat{n}_{i \downarrow}\right)$ \cite{ZGR,Karyn,Gutzwiller}.

We account for the strong repulsion by projecting onto the states with site-population less than or equal to unity which is done by introducing the site-dependent Gutzwiller factors \cite{ZGR}: 
 \begin{eqnarray}
 g_\mathrm{t}^{ij} &=& \sqrt{g_\mathrm{t}^{i}g_\mathrm{t}^{j}}=\sqrt{2\delta^{i}/(1+\delta^{i})}\sqrt{2\delta^{j}/(1+\delta^{j})} \\ \nonumber
 g_\mathrm{s}^{ij} &=& \sqrt{g_\mathrm{s}^{i}g_\mathrm{s}^{j}}=\sqrt{4/(1+\delta^{i})^{2}}\sqrt{4/(1+\delta^{j})^{2}}.
 \end{eqnarray}
where we denote the deviation from half-filling $\delta^{i}=1-\braket{\hat{n}_{i}}$ with $\braket{\hat{n}_{i}}=\sum_{\sigma}\braket{\hat{n}_{i\sigma}}$ representing the total number of particles on site $i$. This is justified by choosing a small curvature of the external trapping potential compared to the hopping amplitude. This procedure is shown to be in good agreement with variational Monte Carlo calculations for the projected $d$-wave state \cite{Paramekanti}.  The renormalized Hamiltonian then takes the form
\begin{align}
\hat{H}&=-t\sum_{\braket{ij} \sigma}\left[g_\mathrm{t}^{ij}\hat{c}_{i \sigma}^{\dagger}\hat{c}_{j \sigma}+h.c.\right] \nonumber\\
&+\sum_{i \sigma} V_{i} \hat{c}_{i \sigma}^{\dagger}\hat{c}_{i \sigma} + J \sum_{\braket{ij}} g_\mathrm{s}^{ij} \vec{S}_{i} \cdot \vec{S}_{j},
\end{align}
where $g_\mathrm{s}^{ij}$ works as to enhance spin-spin correlations since $g_\mathrm{s}^{ij}\geq 1$ and $g_\mathrm{t}^{ij}$ suppresses the tunneling term since $g_\mathrm{t}^{ij}\leq 1$ and goes to zero when the doping goes to zero. This resembles the suppression of the kinetic energy in the Mott insulating state.

In order to solve this many-body Hamiltonian we introduce the Fermi liquid order parameter, $\chi_{ij}$, and the pairing order parameter, $\Delta_{ij}$, with the following averages \cite{ZGR}
\begin{align}
\chi_{ij} &= \frac{3}{4}g_\mathrm{s}^{ij}J \sum_{\sigma}\braket{\hat{c}_{i \sigma}^{\dagger}\hat{c}_{j \sigma}} \nonumber\\
\Delta_{ij}&=\frac{3}{4}g_\mathrm{s}^{ij}J \sum_{\sigma \sigma '}\epsilon_{\sigma \sigma'} \braket{\hat{c}_{i \sigma}\hat{c}_{j \sigma'}},
\end{align}
where $\epsilon_{\uparrow \downarrow}=1=-\epsilon_{\downarrow \uparrow}$ and $\epsilon_{\uparrow \uparrow}=0=\epsilon_{\downarrow \downarrow}$. At a general level, spin fluctuations not too far from half-filling will favor d-wave superconductivity. We then obtain the mean-field Hamiltonian 

\begin{align}
&\hat{H}=-\sum_{\braket{ij} \sigma}\left[\left(tg_\mathrm{t}^{ij} +\frac{\chi_{ji}}{2}\right)\hat{c}_{i \sigma}^{\dagger}\hat{c}_{j \sigma}+h.c.\right]-\mu \sum_{i \sigma}\hat{c}_{i \sigma}^{\dagger}\hat{c}_{i \sigma} \nonumber\\
&+\sum_{i \sigma} V_{i} \hat{c}_{i \sigma}^{\dagger}\hat{c}_{i \sigma} +  \sum_{\braket{ij}}\left[\frac{\Delta_{ij}}{2}\left(\hat{c}_{i \uparrow}^{\dagger}\hat{c}_{j \downarrow}^{\dagger}-\hat{c}_{i \downarrow}^{\dagger}\hat{c}_{j \uparrow}^{\dagger}\right) +h.c.\right],
\end{align}
where we now work in the grand-canonical ensemble and therefore introduce the chemical potential $\mu$ since the number of particles is no longer conserved. We have in the above expression assumed that the potential is sufficiently slowly varying so that $\braket{\hat{c}_{i \uparrow}\hat{c}_{j \downarrow}}=\braket{\hat{c}_{j \uparrow}\hat{c}_{i \downarrow}}$. Furthermore, we have exploited the fact that the additional term to the tunneling amplitude is spin-independent $\braket{\hat{c}_{i \uparrow}^{\dagger}\hat{c}_{j \uparrow}}=\braket{\hat{c}_{i \downarrow}^{\dagger}\hat{c}_{j \downarrow}}$ under our present assumptions. 

Below, we show that the pairing order parameter depends strongly on the position in the trap. It converges to the untrapped case in the centre of the trap where it is maximal, which reflects the emergence of superfluidity induced by the pairing term. We confirm the co-existence of a $d$-wave superfluid domain and a gapless edge region. 

Since the trap breaks the discrete translational invariance, we first find the gap by solving the Bogoliubov-de Gennes (BdG) equations.
We diagonalize the renormalized mean-field Hamiltonian by making a unitary Bogoliubov-Valatin transformation of the creation- and annihilation operators, expanding them on a complete basis of quasi-particle modes annihilated by the fermionic mode operators $\hat{\gamma}_{\eta \sigma}$ \cite{deGennes}

\begin{align}
\hat{c}_{ i \uparrow}&= \sum_{\eta} \left(u_{\eta \uparrow}^{i}\hat{\gamma}_{\eta \uparrow}-v_{\eta \downarrow}^{i *} \hat{\gamma}_{\eta \downarrow}^{\dagger} \right)\nonumber\\
\hat{c}_{ i \downarrow}&= \sum_{\eta}\left( u_{\eta \downarrow}^{i} \hat{\gamma}_{\eta \downarrow}+v_{\eta \uparrow}^{i *} \hat{\gamma}_{\eta \uparrow}^{\dagger}\right).
\end{align}
Here the sums are performed solely over positive energies and $u_{\eta \uparrow}^{i}$ is the amplitude of destroying a quasi-particle with spin up at the $i$'th lattice site. The self-consistent equations of the mean-field parameters take the following form in terms of the 
quasi-particle amplitudes

\begin{align}
\Delta_{ij}&=\frac{3g_\mathrm{s}^{ij}J}{4}\sum_{\eta}\left[ u_{\eta \uparrow}^{i}v_{\eta \uparrow}^{j *} f\left(-E_{\eta \uparrow}\right)- u_{\eta \uparrow}^{j}v_{\eta \uparrow}^{i *} f\left(E_{\eta \uparrow}\right)\right] \nonumber\\
\chi_{ij}&=\frac{3g_\mathrm{s}^{ij}J}{4}\sum_{\eta} \left[ u_{\eta \uparrow}^{i*} u_{\eta \uparrow}^{j} f\left(E_{\eta \uparrow}\right)+ v_{\eta \uparrow}^{i} v_{\eta \uparrow}^{j*} f\left(-E_{\eta \uparrow}\right)\right] \nonumber\\
\delta^{i}&=1-\sum_{\eta}\left[|u_{\eta \uparrow}^{i}|^{2} f(E_{\eta \uparrow})+|v_{\eta \uparrow}^{i}|^{2} f(-E_{\eta \uparrow})\right]
\end{align}
where $f(E_{\eta \uparrow})=\braket{\hat{\gamma}_{\eta \uparrow}^{\dagger}\hat{\gamma}_{\eta \uparrow}}=1/(\exp(E_{\eta \uparrow}/T)+1)$ is the thermal occupation of the quantum state with energy $E_{\eta \uparrow}$. In the above expression the duality of the spin-up and spin-down solutions has been exploited to transform the sum into a sum over positive as well as negative energies for spin-up particles only. 

Assuming a small curvature of the harmonic trapping potential, it is possible to approximate the potential with a constant in the vicinity of each lattice site and replace the eigenstates with plane wave states. Assuming a large number of particles in each cell of constant potential, we define a local Fermi sea. This consists in defining a local chemical potential in each cell $\mu\left(\vec{r}\right) = \mu -V(x,y)$ and letting 
\begin{align}
u_{\eta \uparrow}(\vec{r}) = \frac{u_{\vec{k} \uparrow}\left(\vec{r}\right)e^{-i \vec{k} \cdot \vec{r}}}{\sqrt{N_{L}}}, \ 
v_{\eta \uparrow}(\vec{r}) = \frac{v_{\vec{k} \uparrow}\left(\vec{r}\right)e^{-i \vec{k} \cdot \vec{r}}}{\sqrt{N_{L}}}
\end{align}
 where $u_{\vec{k}}(\vec{r})$ and $v_{\vec{k}}(\vec{r})$ are solutions of the homogenous system with $\mu(\vec{r})=\mu-V(x,y)$ \cite{PhysRevA.73.051602} and $N_L$ is the number of lattice sites. Below, we introduce the local variables
 \begin{align}
 \Delta(x,y)=&\ \Delta_{i}=\frac{1}{N_{N}}\sum_{\hat{a}}\alpha_{\hat{a}}\Delta_{i,i+\hat{a}} \nonumber\\
 \chi(x,y)=&\ \chi_{i}=\frac{1}{N_{N}}\sum_{\hat{a}}\chi_{i,i+\hat{a}}, \nonumber\\
 \delta(x,y)=&\ \delta^i
 \end{align}
where ${\hat{a}}$ denotes the nearest neighbors, $N_{N}$ the number of nearest neighbors, and 
$\alpha_{dx}=\alpha_{-dx}=-\alpha_{dy}=-\alpha_{-dy}=1$.
 
 Within this local density approximation (LDA) the self-consistent equations take the form
\begin{align}
\chi(x,y)=&-\frac{3g_\mathrm{s}(x,y)J }{4 N_{L}} \sum_{\vec{k}}\left(\cos(k_{x}a)+\cos(k_{y}a)\right)\nonumber\\
&\times \tanh\left(\frac{E_{\vec{k}}(x,y) }{2T}\right)\frac{\xi_{\vec{k}}(x,y)}{2E_{\vec{k}(x,y)}} 
\label{chi-LDA}
\end{align}

\begin{align}
\delta(x,y)=\frac{1}{ N_{L}} \sum_{\vec{k}} \tanh\left(\frac{E_{\vec{k}}(x,y) }{2T}\right)\frac{\xi_{\vec{k}}(x,y)}{E_{\vec{k}}(x,y)} 
\end{align}

\begin{align}
\Delta(x,y)=&\frac{3 g_\mathrm{s}(x,y)J}{4 N_{L}}  \sum_{\vec{k}}\left(\cos(k_{x}a)-\cos(k_{y}a)\right) \nonumber\\
&\times \tanh\left(\frac{E_{\vec{k}}(x,y)}{2T}\right) \frac{\Delta_{\vec{k}}(x,y)}{2E_{\vec{k}}(x,y)} ,
\end{align}
with
\begin{align}
E_{\vec{k}}(x,y)= &\sqrt{\xi_{\vec{k}}(x,y)^{2}+\Delta_{\vec{k}}(x,y)^{2}} \nonumber\\
\xi_{\vec{k}}(x,y)= &-(2tg_{t}(x,y)+\chi(x,y))\left[\cos(k_{x}a)+\cos(k_{y}a)\right] \nonumber\\
&-\mu(x,y)\nonumber\\
\Delta_{\vec{k}}(x,y)= &\Delta(x,y) \left[\cos(k_{x}a)-\cos(k_{y}a)\right].
\end{align}

We have neglected the variation in chemical potentials on neighboring lattice sites and approximated the renormalization factors with their value on the $i$'th lattice site, for example $g_\mathrm{s}(x,y)=4/(1+\delta(x,y))^{2}$, which is justified when the curvature of the harmonic potential is small compared to the tunneling amplitude. In the absence of the harmonic trap, these equations and their solutions agree with well-known results; see for example Ref. \cite{Karyn2}.

\begin{figure}
\begin{center}%
\includegraphics[width=240pt,clip=true]{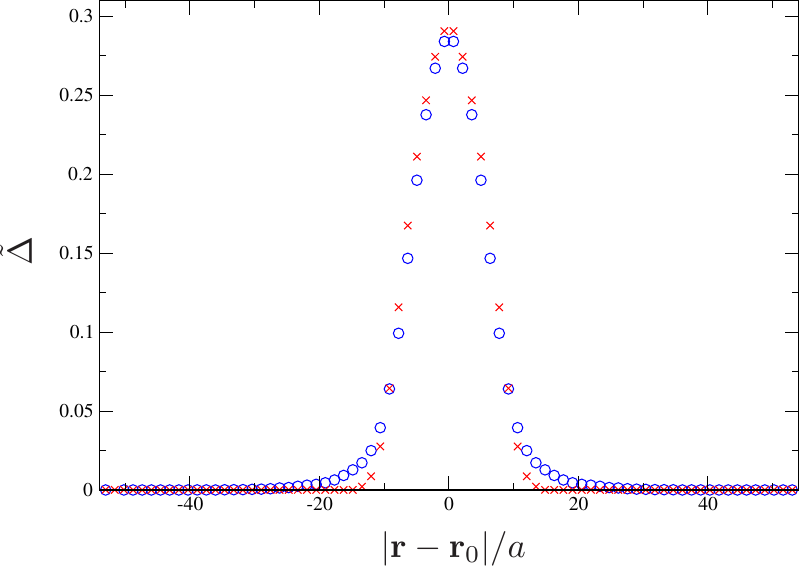}\\
\end{center}
\caption{(Color online) A diagonal cross-section of $\tilde{\Delta}(\mathbf{r})$ of the trapped system with $76\times76$ lattice sites, as a function of the distance from the center of the trap $\mathbf{r}_0$. $t/J = 5$, $\mu/J = -0.1$, and $\tilde{\omega}=\sqrt{ma^{2}/J}\omega = 0.16$. BdG (circles) and LDA (crosses).}%
\label{fig:Delta-trap}%
\end{figure}

We now compare the local density approximation calculation to the BdG calculation of the site-dependent values of $\tilde{\Delta}_{i}$, $\tilde{\chi}_{i}$ and $\braket{\hat{n}_{i}}$; note that the tilde values mean that the mean-field order parameters are expressed in units of $3/4g_\mathrm{s}^{ii}J$. For large interactions, we check that the order parameters are controlled by the Anderson super-exchange $J$. We expect that the critical distance at which $\tilde{\Delta}_{i}$ will be zero is given by the position with a doping corresponding to the critical ``doping'' from the homogenous calculation which lies around $\delta=0.35$ \cite{Karyn2} if one assumes the LDA is correct.  Remarkably, we obtain a very good quantitative agreement between the LDA approximation and the BdG calculation as shown in Figs. 1 and 2. We corroborate the coexistence between $d$-wave superfluidity and normal normal edges in accordance with results from a calculation on a similar system  sufficiently away from half-filling in the weakly interacting limit \cite{Andersen}.  In addition, we find that in the center of the trap, since
the fermion density reaches one, there is a reminiscence of the Mott insulating state which results in a downturn of the Fermi liquid order parameter $\tilde{\chi}_i$.

\begin{figure}
\begin{center}%
\includegraphics[width=240pt,clip=true]{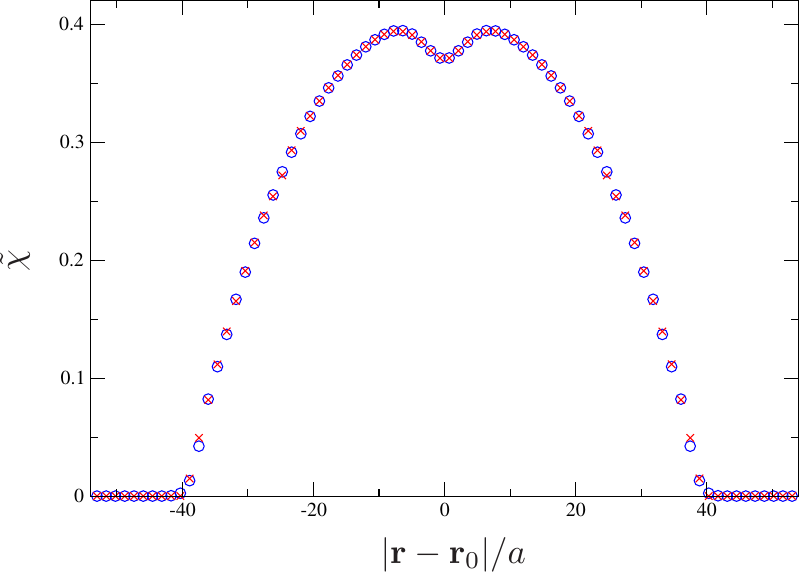}\\
\end{center}
\caption{(Color online) A diagonal cross-section of $\tilde{\chi}(\mathbf{r})$ of the trapped system with $76\times76$ lattice sites, as a function of the distance from the center of the trap $\mathbf{r}_0$. $t/J = 5$, $\mu/J = -0.1$, and $\tilde{\omega}=\sqrt{ma^{2}/J}\omega = 0.16$. BdG (circles) and LDA (crosses).}%
\label{fig:chi-trap}%
\end{figure}

In the following we consider the variation of the doping close to the boundaries of the density profile defined as the circle in the $xy$-plane with radius $R$ where $\delta(x,y)$ reaches unity in the local density approximation. This circle is only well-defined in the continuous system but becomes sufficiently well-defined for large lattices. In the vicinity of the density boundaries the pairing order parameter has vanished, $\chi(x,y)$ is so small that we ignore it and at the boundaries the chemical potential $\mu(R)$ is equal to $-4t$. This means that all the terms in the expression for $\delta(R)$ contribute with a positive sign since $\xi_{\vec{k}}(R)=-2t\left(\cos(k_{x}a)+\cos(k_{y}a)\right)+4t>0$ for all quasi-momenta and $\delta(R)$ is therefore maximal and unity. In a small region close to $\delta(R)=1$ the terms in the sum which contribute to a lowering of the doping concentration will form a circle in $\vec{k}$-space. Hence, we make a Taylor expansion of the cosine terms in $\vec{k}$-space at site $i$ with coordinates $(x,y)$ to find the radius $ka$ corresponding to the change of sign in $\xi_{{\vec k}}(x,y)$: $-2tg_{t}(x,y)(\cos(k_{x}a)+\cos(k_{y}a)) -\mu(x,y)\approx -2tg_{t}(x,y)(2-1/2\left[(k_{x}a)^{2}+(k_{y}a)^{2}\right])-\mu(x,y)=-2tg_{t}(x,y)(2-1/2 (ka)^{2})-\mu(x,y)=0$ which corresponds to a radius of $(ka)^{2}=\frac{\mu(x,y)}{tg_{t}(x,y)}+4$. The doping will then be given by
\begin{align}
\delta(x,y)&=\frac{1}{N_{L}}\sum_{\vec{k}}\hbox{sign}(\xi_{\vec{k}}(x,y)) \nonumber\\
&=\frac{4 \pi^2-2\pi (ka)^{2}}{4 \pi^2}  =1-\frac{\frac{\mu(x,y)}{tg_{t}(x,y)}+4}{2\pi} \nonumber\\
&=1-\frac{2}{\pi}-\frac{\mu(x,y)}{4\pi t \delta(x,y)}-\frac{\mu(x,y)}{4\pi t }  \nonumber\\
&=\frac{1}{2}-\frac{1}{\pi}\left[\frac{\mu(x,y)}{8t}+1\right] \nonumber\\
&+\frac{\sqrt{\left(1-\frac{2}{\pi}\left(\frac{\mu(x,y)}{8t}+1\right)\right)^{2}-\frac{\mu(x,y)}{\pi t}}}{2},
\end{align}
with $\mu(x,y)=\mu-V(x,y)$ the local chemical potential when realizing that one must subtract the area corresponding to the negative contribution twice. 

\begin{figure}
\begin{center}%
\includegraphics[width=240pt,clip=true]{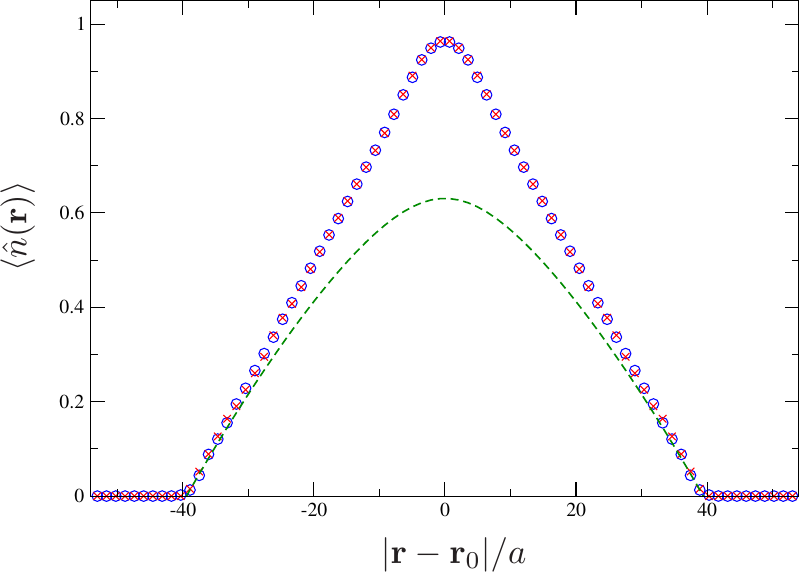}\\
\end{center}
\caption{(Color online) A diagonal cross-section of $\langle \hat{n}(\mathbf{r}) \rangle$ of the trapped system with $76\times76$ lattice sites, as a function of the distance from the center of the trap $\mathbf{r}_0$. $t/J = 5$, $\mu/J = -0.1$, and $\tilde{\omega}=\sqrt{ma^{2}/J}\omega = 0.16$. BdG (circles),
LDA (crosses) and analytical (dashed).}%
\label{fig:n-trap}%
\end{figure}

One can use the expression for the chemical potential at the boundaries $\mu(R)=\mu-1/2m\omega^{2}R^{2}=-4t$ to obtain the associated radius $R=\sqrt{2\left(4t+\mu\right)/\left(m\omega^{2}\right)}$,
which corresponds to a radius of $R=39.5a$ when $t/J=5$, $T/J=0$, $\mu/J=-0.1$, and $\tilde{\omega}=0.16$ which is shown to be correct in Fig. \ref{fig:n-trap}.

Due to the isotropy of the harmonic potential the expression for the doping is assumed to have no angular dependence as in the continuous case and the  expression for the density can be Taylor expanded around $R$ to first order in $x$ to give
$n(x,0)=-0.024(x/a-39.5)$.

A qualitative explanation of $\tilde{\Delta}_{i}$ is given at finite temperature. We observe that the height and the width of $\tilde{\Delta}_{i}$ decreases with increasing temperature whereas the shape does not change significantly. 

To summarize, we have shown that a harmonic trapping potential can simulate the effect of inhomogeneous doping allowing to stabilize a novel $d$-wave superfluid phase for fermions with gapless edge states. In the context of strongly repulsive fermions, we have found that at the center of the trap there is a reminiscence of the Mott insulating state which is revealed by a downturn of the Fermi liquid order parameter whereas the $d$-wave gap is maximal. The proximity to the Mott insulating state at the center of the trap should affect the double
occupancy. The $d$-wave symmetry of the pairing can be detected through the bunching which is maximal along the $x$-and $y$-directions and minimal along $x=\pm y$. The pairing should be maximal around the Fermi surface. Phase sensitive measurements could also be performed \cite{Aspect}. We found a linear profile of the fermion density close to the boundaries. When decreasing the density, one might observe magnetic real-space shell structures competing with the superfluid phase \cite{Andersen}. For dipolar fermions, $d$-wave bond order solidity might also occur at half-filling \cite{Zhao}.

We are grateful to N. G. Nygaard for valuable discussions concerning the density profile. This work is supported by DARPA W911NF-10-1-0206 and NSF DMR0803200.


\begin{thebibliography}{99}

\bibitem{Zoller}
D. Jaksch and P. Zoller, Ann. Phys. {\bf 35}, 52 (2005).

\bibitem{Bloch}
I. Bloch, J. Dalibard, and W. Zwerger, Rev. Mod. Phys. {\bf 80}, 885 (2008).

\bibitem{Georges}
A. Georges, in {\it Ultra-cold Fermi gases}, edited by M. Inguscio, W. Ketterle and C. Salomon (Italian physical
society, 2007), p. 477, arXiv:cond-mat/0702122.

\bibitem{Greiner}
M. Greiner, O. Mandel, T. Esslinger, T. W. H\" ansch, and I. Bloch, Nature (London) {\bf 415}, 39 (2002).

\bibitem{Trotzky}
S. Trotzky {\it et al.}, Nature Phys. {\bf 6}, 998-1004 (2010).

\bibitem{Kohl}
M. K\" ohl, H. Mortiz, T. St\" oferle, K. G\" unter, and T. Esslinger, Phys. Rev. Lett. {\bf 94}, 080403 (2005).

\bibitem{Jin}
J. T. Stewart, J. P. Gaebler, and D. S. Jin, Nature {\bf 454}, 744 (2008).

\bibitem{Greiner2}
J. Simon, W. Bakr, R. Ma, M. E. Tai, P. M. Preiss, and M. Greiner, Nature {\bf 472}, 307 (2011).

\bibitem{Folling}
S. F\" olling {\it et al.}, Nature (London) {\bf 434}, 481 (2005).

\bibitem{Esslinger}
R. J\" ordens {\it et al.}, Nature (London) {\bf 455}, 204-207 (2008).

\bibitem{Bloch2}
U. Schneider {\it et al.}, Science {\bf 322}, 1520-1525 (2008).

\bibitem{Wolfgang}
J. K. Chin, D. E. Miller, Y. Liu, C. Stan, W. Setiawan, C. Sanner, K. Xu, and W. Ketterle, Nature (London) {\bf 443}, 961 (2006).

\bibitem{Jordens}
R. J\" ordens {\it et al.}, Phys. Rev. Lett. {\bf 104}, 180401 (2010).

\bibitem{Hofstetter}
W. Hofstetter, J. I. Cirac, P. Zoller, E. Demler and M. D. Lukin, Phys. Rev. Lett. {\bf 89}, 220407 (2002).

\bibitem{Anderson}
P. W. Anderson, Science {\bf 235}, 1196 (1987).

\bibitem{Lee}
P. A. Lee, N. Nagaosa, and X. G. Wen, Rev. Mod. Phys. {\bf 78}, 17 (2006).

\bibitem{Ogata}
M. Ogata and H. Fukuyama, Rep. Prog. Phys. {\bf 71}, 036501 (2008).

\bibitem{Karyn}
Karyn Le Hur and T. Maurice Rice, Ann. Phys. {\bf 324}, 1452 (2009).

\bibitem{Wolfgang2}
C. Sanner {\it et al.}, Phys. Rev. Lett. {\bf 106}, 010402 (2011).

\bibitem{Andersen} B. M. Andersen and G. M. Bruun, Phys. Rev. A {\bf 76}, 041602 (2007).

\bibitem{ZGR} F. C. Zhang, C. Gros, T. M. Rice, and H. Shiba, Supercond. Sci. Technol. {\bf 1}, 36 (1988).

\bibitem{Anderson2}
P. W. Anderson {\it et al.}, J. Phys.: Condens. Matter {\bf 16}, R755-769 (2004).

\bibitem{Karyn2} K. Le Hur, C.-H. Chung, and I. Paul, Phys. Rev. B {\bf 84}, 024526 (2011). 

\bibitem{Gutzwiller}
M. C. Gutzwiller, Phys. Rev. {\bf 137}, A1726 (1965).

\bibitem{Paramekanti}
A. Paramekanti, M. Randeria and N. Trivedi, Phys. Rev. Lett. {\bf 87}, 217002 (2001).

\bibitem{deGennes} P. G. de Gennes: \emph{Superconductivity of Metals and Alloys}, (W. A. Benjamin, Inc., New York, 1966).

\bibitem{PhysRevA.73.051602} T. N. De Silva and E. J. Mueller, Phys. Rev. A {\bf 73}, 051602(R) (2006).

\bibitem{Aspect}
T. Kitagawa, A. Aspect, M. Greiner and E. Demler, Phys. Rev. Lett. {\bf 106}, 115302 (2011).

\bibitem{Zhao}
S. G. Bonghale {\it et al.} arXiv:1111.2873.

\end{thebibliography}
\end{document}